# The Heisenberg versus the Schrödinger picture in quantum field theory

by

Dan Solomon
Rauland-Borg Corporation
3450 W. Oakton
Skokie, IL 60077
USA

Phone: 847-324-8337
Email: **dan.solomon@rauland.com**

**Abstract**

The Heisenberg picture and Schrödinger picture are supposed to be equivalent representations of quantum mechanics. However this idea has been challenged by P.A.M. Dirac [4]. Also, it has been recently shown by A. J. Faria et al [3] that this is not necessarily the case. In this article a simple problem will be worked out in quantum field theory in which the Heisenberg picture and Schrödinger picture give different results.

**I. Introduction**

The Heisenberg picture and Schrödinger picture are supposed to be equivalent representations of quantum theory [1][2]. However A.J. Faria et al[3] have recently presented an example in non-relativistic quantum theory where they claim that the two pictures yield different results. Previously P.A.M. Dirac [4] has suggested that the two pictures are not equivalent. In order to further investigate this problem we will examine quantum field theory in both the Heisenberg and Schrödinger pictures and will show that, even though they are formally equivalent, they yield different results when an actual problem is worked out.

In quantum field theory a quantum system, at a given point in time, is specified by the state vector $|\Omega\rangle$ and field operator $\hat{\psi}(\vec{x})$. We will write this as the pair $(|\Omega\rangle, \psi)$. Let the state vector $|\Omega\rangle$ and the field operator $\hat{\psi}(\vec{x})$ be defined at some initial point in time, say t=0. This may be taken as the initial conditions of the quantum system. Now there are two ways to handle the time evolution of the system. In the Schrödinger picture it is assumed that field operator $\hat{\psi}(\vec{x})$ is constant in time and the time dependence of the system goes with the state vector $|\Omega(t)\rangle$. In the Heisenberg picture the time dependence is assigned to the field operator $\hat{\psi}(\vec{x},t)$ and the state vector $|\Omega\rangle$ remains constant in time. Both pictures are supposed to give equivalent results in that the expectation values of operators are the same. Note that at the initial time, t=0, both pictures are identical. Therefore the time independent Schrödinger field operator $\hat{\psi}(\vec{x})$ is equal to $\hat{\psi}(\vec{x},0)$, which is the time dependent Heisenberg field operator at t=0.

Similarly, the time independent Heisenberg state vector $|\Omega\rangle$ equals $|\Omega(0)\rangle$, which is the time dependent Schrödinger state vector $|\Omega(t)\rangle$ at t=0. For example, let the initial state of the system, at t=0, be represented by the pair $(|\Omega(0)\rangle, \psi(\vec{x}, 0))$. In the Heisenberg picture this initial state evolves into $(|\Omega(0)\rangle, \psi(\vec{x}, t))$. In the Schrödinger picture the state evolves into $(|\Omega(t)\rangle, \psi(\vec{x}, 0))$.

From Chapter 9 of Greiner et al [5] the Dirac Hamiltonian in the presence of a classical electromagnetic field is given by,

$$\hat{H}(\hat{\psi}) = \frac{1}{2}\int \left[\hat{\psi}^\dagger, H\hat{\psi}\right] d\vec{x} - \xi_r \tag{1}$$

where,

$$H = H_0 - q\vec{\alpha}\cdot\vec{A} + qA_0 \tag{2}$$

and

$$H_0 = -i\vec{\alpha}\cdot\vec{\nabla} + \beta m \tag{3}$$

and $\xi_r$ is a renormalization constant. In the above expression $(A_0, \vec{A})$ is the classical electric potential and is taken to be an unquantized, real valued quantity. Also q and m are the charge and mass of the electron, respectively, and the 4x4 matrices $\vec{\alpha}$ and $\beta$ are defined in [5]. Note that in the above equations we use $\hbar = c = 1$. The term in the brackets in (1) is defined by the following expression,

$$\left[\hat{\psi}^\dagger, H\hat{\psi}\right] = \hat{\psi}^\dagger(H\hat{\psi}) - (H\hat{\psi})^T(\hat{\psi}^\dagger)^T = \hat{\psi}_\alpha^\dagger(H_{\alpha\beta}\hat{\psi}_\beta) - (H_{\alpha\beta}\hat{\psi}_\beta)\hat{\psi}_\alpha^\dagger \tag{4}$$

where the summation is over the spin indices.

Equation (1) can also be rewritten as,

$$\hat{H}(\hat{\psi}) = \hat{H}_0(\hat{\psi}) - \int \hat{\vec{J}}(\hat{\psi}) \cdot \vec{A}(\vec{x},t) d\vec{x} + \int \hat{\rho}(\hat{\psi}) A_0(\vec{x},t) d\vec{x} \tag{5}$$

where

$$\hat{H}_0(\hat{\psi}) = \frac{1}{2}\int\left[\hat{\psi}^\dagger, H_0\hat{\psi}\right]d\vec{x} - \xi_r \tag{6}$$

and

$$\hat{\vec{J}}(\hat{\psi}) = \frac{q}{2}\left[\hat{\psi}^\dagger, \vec{\alpha}\hat{\psi}\right] \text{ and } \hat{\rho}(\hat{\psi}) = \frac{q}{2}\left[\hat{\psi}^\dagger, \hat{\psi}\right] \tag{7}$$

$\hat{H}_0(\hat{\psi})$ is the free field Hamiltonian operator, $\hat{\vec{J}}(\hat{\psi})$ is the current operator, and $\hat{\rho}(\hat{\psi})$ is the charge operator.

In the Heisenberg picture the evolution of the field operator is given by,

$$\frac{\partial \hat{\psi}(\vec{x},t)}{\partial t} = i\left[\hat{H}(\hat{\psi}(\vec{x},t)), \hat{\psi}(\vec{x},t)\right] \tag{8}$$

The field operator obeys the equal time anti-commutator relationship,

$$\hat{\psi}^\dagger_\alpha(\vec{x},t)\hat{\psi}_\beta(\vec{x}',t) + \hat{\psi}_\beta(\vec{x}',t)\hat{\psi}^\dagger_\alpha(\vec{x},t) = \delta_{\alpha\beta}\delta^{(3)}(\vec{x}-\vec{x}') \tag{9}$$

with all other equal time anti-commutators being equal to zero. It is shown by Greiner (see Chapt. 9 of [5] or Section 8 of [1]) that when these are used in (8) we obtain,

$$i\frac{\partial \hat{\psi}(\vec{x},t)}{\partial t} = H\hat{\psi}(\vec{x},t) \tag{10}$$

To convert to the Schrödinger picture define the operator $\hat{U}(t)$ which satisfies the differential equation,

$$i\frac{\partial \hat{U}(t)}{\partial t} = \hat{H}(\hat{\psi}(\vec{x},0))\hat{U}(t); \quad -i\frac{\partial \hat{U}^\dagger(t)}{\partial t} = \hat{U}^\dagger(t)\hat{H}(\hat{\psi}(\vec{x},0)) \tag{11}$$

where $\hat{U}(t)$ is subject to the initial condition $\hat{U}(0)=1$. From the above expressions we have that $\partial\left(\hat{U}^\dagger(t)\hat{U}(t)\right)/\partial t = 0$. From this, and the initial condition on $\hat{U}(t)$ we obtain $\hat{U}^\dagger(t)\hat{U}(t)=1$ which yields,

$$\hat{U}^\dagger(t) = \hat{U}^{-1}(t) \tag{12}$$

The time dependent Schrödinger state vector is defined by,

$$|\Omega(t)\rangle = \hat{U}(t)|\Omega(0)\rangle \text{ and } \langle\Omega(t)| = \langle\Omega(0)|\hat{U}^\dagger(t) \tag{13}$$

Use this and (11) to show that $|\Omega(t)\rangle$ satisfies,

$$i\frac{\partial|\Omega(t)\rangle}{\partial t} = \hat{H}(\hat{\psi}(\vec{x},0))|\Omega(t)\rangle \tag{14}$$

## II. The Vacuum state

In this section we will develop some of the tools of quantum field theory in the Schrödinger picture. In particular we will reach some conclusions concerning the expectation value of the operator $\hat{H}_0$ in the Schrödinger picture.

Define $\phi_n^{(0)}(\vec{x})$ as being basis states solutions of the following equation,

$$H_0\phi_n^{(0)}(\vec{x}) = \lambda_n E_n \phi_n^{(0)}(\vec{x}) \tag{15}$$

where the energy eigenvalues $\lambda_n E_n$ and can be expressed by

$$E_n = +\sqrt{\vec{p}_n^2 + m^2}, \quad \lambda_n = \begin{cases} +1 \text{ for a positive energy state} \\ -1 \text{ for a negative energy state} \end{cases} \tag{16}$$

and where $\vec{p}_n$ is the momentum of the state n.

The $\phi_n^{(0)}(\vec{x})$ can be expressed by,

$$\phi_n^{(0)}(\vec{x}) = u_n e^{i\vec{p}\cdot\vec{x}} \tag{17}$$

where $u_n$ is a constant 4-spinor. The $\phi_n^{(0)}(\vec{x})$ form a complete orthonormal basis in Hilbert space and satisfy

$$\int \phi_n^{(0)\dagger}(\vec{x}) \phi_m^{(0)}(\vec{x}) d\vec{x} = \delta_{mn} \tag{18}$$

Define the index 'n' so that it is positive for positive energy states and negative for negative energy states. The time independent Schrödinger field operator can, then, be defined as follows,

$$\hat{\psi}_S(\vec{x}) = \sum_{n>0} \left( \hat{b}_n \phi_n^{(0)}(\vec{x}) + \hat{d}_n^\dagger \phi_{-n}^{(0)}(\vec{x}) \right); \quad \hat{\psi}_S^\dagger(\vec{x}) = \sum_{n>0} \left( \hat{b}_n^\dagger \phi_n^{(0)\dagger}(\vec{x}) + \hat{d}_n \phi_{-n}^{(0)\dagger}(\vec{x}) \right) \tag{19}$$

where the $\hat{b}_n$ ($\hat{b}_n^\dagger$) are the destruction(creation) operators for an electron associated with the state $\phi_n^{(0)}(\vec{x})$ and the $\hat{d}_n$ ($\hat{d}_n^\dagger$) are the destruction(creation) operators for a positron associated with the state $\phi_{-n}^{(0)}(\vec{x})$. They satisfy the anticommutator relationships,

$$\{\hat{d}_j, \hat{d}_k^\dagger\} = \delta_{jk}; \quad \{\hat{b}_j, \hat{b}_k^\dagger\} = \delta_{jk}; \text{ all other anti-commutators are zero} \tag{20}$$

The vacuum state $|0\rangle$ is defined by,

$$\hat{d}_j |0\rangle = \hat{b}_j |0\rangle = 0 \text{ and } \langle 0| \hat{d}_j^\dagger = \langle 0| \hat{b}_j^\dagger = 0 \tag{21}$$

Using the above results the free field Hamiltonian operator becomes,

$$\hat{H}_0(\hat{\psi}_S) = \frac{1}{2} \sum_{n>0} \left( \lambda_n E_n \left( \hat{b}_n^\dagger \hat{b}_n - \hat{b}_n \hat{b}_n^\dagger \right) + \lambda_{-n} E_{-n} \left( \hat{d}_n \hat{d}_n^\dagger - \hat{d}_n^\dagger \hat{d}_n \right) \right) - \xi_r \tag{22}$$

Use (20) in the above to yield,

$$\hat{H}_0(\hat{\psi}_S) = \sum_{n>0} \left( E_n \hat{b}_n^\dagger \hat{b}_n + E_n \hat{d}_n^\dagger \hat{d}_n \right) - E_{vac} - \xi_r \tag{23}$$

where we have used $E_n = E_{-n}$, $\lambda_n = 1$, and $\lambda_{-n} = -1$ for $n > 0$ and,

$$E_{vac} = \sum_{n>0} E_n \qquad (24)$$

Next define $\xi_r = -E_{vac}$ to obtain,

$$\hat{H}_0(\hat{\psi}_S) = \sum_{n>0} E_n \left( \hat{b}_n^\dagger \hat{b}_n + \hat{d}_n^\dagger \hat{d}_n \right) \qquad (25)$$

The vacuum state $|0\rangle$ is an eigenstate of the free field Hamiltonian $\hat{H}_0(\hat{\psi}_S)$ with eigenvalue $\varepsilon(|0\rangle) = 0$, i.e.,

$$\hat{H}_0(\hat{\psi}_S)|0\rangle = 0 \qquad (26)$$

New eigenstates, $|n\rangle$ can be produced by operating on $|0\rangle$ with various combinations of the electron and positron creation operators $\hat{b}_n^\dagger$ and $\hat{d}_n^\dagger$, respectively. Since electrons and positrons are particles with positive energy the new states $|n\rangle$ are positive energy states with respect to the vacuum. The total set of eigenstates $|n\rangle$ (which includes the vacuum state $|0\rangle$) form an orthonormal basis that satisfies the following relationships,

$$\hat{H}_0(\hat{\psi}_S)|n\rangle = \varepsilon(|n\rangle)|n\rangle \text{ where } \varepsilon(|n\rangle) > \varepsilon(|0\rangle) = 0 \text{ for } |n\rangle \neq |0\rangle \qquad (27)$$

and

$$\langle n|m\rangle = \delta_{mn} \qquad (28)$$

Any arbitrary state $|\Omega\rangle$ can be expanded in terms of these basis states,

$$|\Omega\rangle = \sum_n c_n |n\rangle \qquad (29)$$

where $c_n$ are the expansion coefficients. Use the above to obtain,

$$\langle \Omega | \hat{H}_0 (\hat{\psi}_S) | \Omega \rangle = \sum_n |c_n|^2 \, \varepsilon(|n\rangle) \tag{30}$$

Since the $\varepsilon(|n\rangle)$ are all non-negative we obtain the relationship,

$$\langle \Omega | \hat{H}_0 (\hat{\psi}_S) | \Omega \rangle \geq 0 \text{ for all } |\Omega\rangle \tag{31}$$

### III. Comparing Expectation values

In this section we will compare the expectation of the operator $\hat{H}_0(\hat{\psi})$ in the Heisenberg and Schrödinger pictures. For the quantum system $(|\Omega\rangle, \hat{\psi})$ the expectation value of $\hat{H}_0$ is given by $\langle \Omega | \hat{H}_0(\hat{\psi}) | \Omega \rangle$. Now consider a system whose initial state, at t=0, is $(|\Omega(0)\rangle, \hat{\psi}_S(\vec{x}))$ and let the electric potential at this time be zero. Now apply an electric potential for some period of time and then remove it at time $t_1 > 0$ so that,

$$(A_0, \vec{A}) = 0 \text{ for } t < 0; \quad (A_0, \vec{A}) \neq 0 \text{ for } 0 \leq t \leq t_1; \quad (A_0, \vec{A}) = 0 \text{ for } t > t_1 \tag{32}$$

We want to determine the quantity $\langle \Omega | \hat{H}_0(\hat{\psi}) | \Omega \rangle$ at some final time $t_f > t_1$. In the Heisenberg picture the state evolves into $(|\Omega(0)\rangle, \hat{\psi}(\vec{x}, t_f))$ where $\hat{\psi}(\vec{x}, t)$ satisfies Eq. (10) with the initial condition $\hat{\psi}(\vec{x}, 0) = \hat{\psi}_S(\vec{x})$. In the Schrödinger picture the state evolves into $(|\Omega(t_f)\rangle, \hat{\psi}_S(\vec{x}))$ where $|\Omega(t)\rangle$ satisfies Eq. (14). Therefore, in the Heisenberg picture, at the final time $t_f$, the expectation value of $\hat{H}_0(\hat{\psi})$ is given by,

$$H_{0,eH}(t_f) = \langle \Omega(0) | \hat{H}_0(\hat{\psi}(\vec{x}, t_f)) | \Omega(0) \rangle \tag{33}$$

and in the Schrödinger picture the expectation value is given by,

$$H_{0,eS}(t_f) = \langle \Omega(t_f) | \hat{H}_0(\hat{\psi}_S) | \Omega(t_f) \rangle \tag{34}$$

If the two pictures are equivalent then we should have,

$$H_{0,eH}(t) = H_{0,eS}(t) \text{ for all t} \tag{35}$$

This relationship is obviously true at the initial time $t = 0$. In the Appendix a formal proof is presented which shows that this relationship is true for all time. However, when the above problem is actually worked out it is shown that the above relationship does not necessarily hold at the time $t_f$.

Now in general it is not possible to find a solution to Eq. (10) for most non-zero electrical potentials. However we will consider an electric potential for which an exact solution can be calculated. During the time interval $0 \le t \le t_1$ the electric potential is non-zero. Let it be given by,

$$(A_0, \vec{A}) = \left(\frac{\partial \chi}{\partial t}, -\vec{\nabla}\chi\right); \quad 0 \le t \le t_1 \tag{36}$$

where $\chi(\vec{x}, t)$ is an arbitrary real valued function that satisfies the following initial condition at t=0,

$$\chi(\vec{x}, 0) = 0; \quad \left.\frac{\partial \chi(\vec{x}, t)}{\partial t}\right|_{t=0} = 0 \tag{37}$$

Refining to Eqs. (10), (32), and (36) we obtain,

$$i\frac{\partial \hat{\psi}(\vec{x}, t)}{\partial t} = \left(H_0 + q\vec{\alpha} \cdot \vec{\nabla}\chi + q\frac{\partial \chi}{\partial t}\right)\hat{\psi}(\vec{x}, t) \text{ for } 0 \le t \le t_1 \tag{38}$$

and,

$$i\frac{\partial \hat{\psi}(\vec{x}, t)}{\partial t} = H_0 \hat{\psi}(\vec{x}, t) \text{ for } t > t_1 \tag{39}$$

Since the time derivative is to the first order the boundary condition at $t = t_1$ is,

$$\hat{\psi}(\vec{x}, t_1 + \delta) \underset{\delta \to 0}{=} \hat{\psi}(\vec{x}, t_1 - \delta) \tag{40}$$

The solution to (38) is,

$$\hat{\psi}(\vec{x},t) = e^{-iq\chi(\vec{x},t)} e^{-iH_0 t} \hat{\psi}_S(\vec{x}) \text{ for } 0 \leq t \leq t_1 \tag{41}$$

where we have used the initial condition $\hat{\psi}(\vec{x},0) = \hat{\psi}_S(\vec{x})$ as well as (37). The solution to (39) is,

$$\hat{\psi}(\vec{x},t_f) = e^{-iH_0(t_f - t_1)} \hat{\psi}(\vec{x},t_1) \text{ for } t_f > t_1 \tag{42}$$

Using the boundary conditions (40) we obtain,

$$\hat{\psi}(\vec{x},t_f) = e^{-iH_0(t_f - t_1)} e^{-iq\chi(\vec{x},t_1)} e^{-iH_0 t_1} \hat{\psi}_S(\vec{x}) \tag{43}$$

Rewrite (43) as,

$$\hat{\psi}(\vec{x},t_f) = e^{-iH_0(t_f - t_1)} e^{-iq\chi(\vec{x},t_1)} \hat{\psi}_0(\vec{x},t_1) \tag{44}$$

where,

$$\hat{\psi}_0(\vec{x},t) = e^{-iH_0 t} \hat{\psi}_S(\vec{x}) \tag{45}$$

Note that $\hat{\psi}_0(\vec{x},t)$ is the field operator that $\hat{\psi}_S(\vec{x})$ would evolve into if the electric potential was zero. Use these results in (6) to yield,

$$\hat{H}_0(\hat{\psi}(\vec{x},t_f)) = \frac{1}{2}\int \left[ \hat{\psi}_0^\dagger(\vec{x},t_1) e^{+iq\chi(\vec{x},t_1)}, H_0 e^{-iq\chi(\vec{x},t_1)} \hat{\psi}_0(\vec{x},t_1) \right] d\vec{x} - \xi_r \tag{46}$$

Use the following result,

$$H_0 e^{-iq\chi} \hat{\psi}_0(\vec{x},t) = e^{-iq\chi}\left(-q\vec{\alpha}\cdot\vec{\nabla}\chi + H_0\right) \hat{\psi}_0(\vec{x},t) \tag{47}$$

in (46) to obtain,

$$\hat{H}_0(\hat{\psi}(\vec{x},t_f)) = \frac{1}{2}\int \left[ \hat{\psi}_0^\dagger(\vec{x},t_1), \left(-q\vec{\alpha}\cdot\vec{\nabla}\chi(\vec{x},t_1) + H_0\right) \hat{\psi}_0(\vec{x},t_1) \right] d\vec{x} - \xi_r \tag{48}$$

Use (7) in the above to obtain,

$$\hat{H}_0\left(\hat{\psi}(\vec{x},t_f)\right) = \hat{H}_0\left(\hat{\psi}_0(\vec{x},t_1)\right) - \int\left(\hat{\vec{J}}\left(\hat{\psi}_0(\vec{x},t_1)\right)\cdot\vec{\nabla}\chi(\vec{x},t_1)\right)d\vec{x} \quad (49)$$

Next sandwich the above expression between $\langle\Omega(0)|$ and $|\Omega(0)\rangle$ and refer to (33) to obtain,

$$\begin{aligned}H_{0,eH}(t_f) &= \langle\Omega(0)|\hat{H}_0\left(\hat{\psi}_0(\vec{x},t_1)\right)|\Omega(0)\rangle \\ &\quad - \int\left(\langle\Omega(0)|\hat{\vec{J}}\left(\hat{\psi}_0(\vec{x},t_1)\right)|\Omega(0)\rangle\cdot\vec{\nabla}\chi(\vec{x},t_1)\right)d\vec{x}\end{aligned} \quad (50)$$

Define the current expectation value by

$$\vec{J}_{0,e}(\vec{x},t) \equiv \langle\Omega(0)|\hat{\vec{J}}\left(\hat{\psi}_0(\vec{x},t)\right)|\Omega(0)\rangle \quad (51)$$

The quantity $\vec{J}_{0,e}(\vec{x},t)$ is the current that would exist at time t if the electric potential was zero during the interval from 0 to t. Use this definition in (50) and assume reasonable boundary conditions and integrate by parts to obtain,

$$H_{0,eH}(t_f) = \langle\Omega(0)|\hat{H}_0\left(\hat{\psi}_0(\vec{x},t_1)\right)|\Omega(0)\rangle + \int\left(\chi(\vec{x},t_1)\vec{\nabla}\cdot\vec{J}_{0,e}(\vec{x},t_1)\right)d\vec{x} \quad (52)$$

In the above expression examine the quantities on the right of the equals sign. Note that $\langle\Omega(0)|\hat{H}_0\left(\hat{\psi}_0(\vec{x},t_1)\right)|\Omega(0)\rangle$ and $\vec{J}_{0,e}(\vec{x},t_1)$ are independent of $\chi(\vec{x},t_1)$. This means that $\chi(\vec{x},t_1)$ can be varied in an arbitrary manner without affecting either of these quantities. Therefore, if $\vec{\nabla}\cdot\vec{J}_{0,e}(\vec{x},t_1)$ is nonzero, it is always possible to specify a $\chi(\vec{x},t_1)$ such that the quantity on the left, $H_{0,eH}$, is a negative number. For example, let $\chi(\vec{x},t_1) = -f\vec{\nabla}\cdot\vec{J}_{0,e}(\vec{x},t_1)$ where f is a real number. Use this in (52) to obtain,

$$H_{0,eH}(t_f) = \langle\Omega(0)|\hat{H}_0\left(\hat{\psi}_0(\vec{x},t_1)\right)|\Omega(0)\rangle - f\int\left(\vec{\nabla}\cdot\vec{J}_{0,e}(\vec{x},t_1)\right)^2 d\vec{x} \quad (53)$$

This quantity will be negative for large enough f since $\langle\Omega(0)|\hat{H}_0(\hat{\psi}_0(\vec{x},t_1))|\Omega(0)\rangle$ and $\vec{\nabla}\cdot\vec{J}_{0,e}(\vec{x},t_1)$ do not vary as f is increased. This result depends on $\vec{\nabla}\cdot\vec{J}_{0,e}(\vec{x},t_1)$ being non-zero. How do we know that this is the case? If quantum mechanics is an accurate model of the real world then there must exist quantum states where $\vec{\nabla}\cdot\vec{J}_{0,e}(\vec{x},t_1)$ is non-zero because there are many examples in the real world where this is the case. For example, $\vec{\nabla}\cdot\vec{J}_{0,e}(\vec{x},t_1)$ will be non-zero for a localized electron wave packet traveling with a non-zero velocity. So that by properly preparing the initial state $|\Omega(0)\rangle$ we can always ensure that $\vec{\nabla}\cdot\vec{J}_{0,e}(\vec{x},t_1)$ will be non-zero.

Now refer back to equation (35). From (53) we see that the left hand side can be negative. However it can be seen by examining (31) and (34) that $H_{0,eS}(t)\geq 0$ which means that the right side of (35) must be non-negative. Therefore the relationship given by (35) does not hold in this case and the Heisenberg picture and Schrödinger picture gives different results for the expectation value in question.

Thus there appears to be a mathematical inconsistency in quantum field theory. On one hand it is possible to show, formally, that the Heisenberg picture and Schrödinger picture are equivalent. On the other hand when an actual problem is worked out we see that they given different results. The reason for this is that in the Schrödinger picture the quantity $\langle\Omega|\hat{H}_0(\hat{\psi}_S(\vec{x}))|\Omega\rangle$ must always be non-negative. It is not possible to find a state vector $|\Omega\rangle$ where $\langle\Omega|\hat{H}_0(\hat{\psi}_S(\vec{x}))|\Omega\rangle$ is negative. In order to eliminate this inconsistency it is necessary to redefine the vacuum state $|0\rangle$ so that eigenstates $|n\rangle$ exist where the $\varepsilon_0(|n\rangle) < 0$. This will allow for the existence of states $|\Omega\rangle$ where $\langle \Omega | \hat{H}_0(\hat{\psi}_S(\vec{x})) | \Omega \rangle$ is negative. The way this can be achieved is examined in [6] and [7].

## V. Summary and Conclusion

It is generally assumed that the Heisenberg and Schrödinger pictures are equivalent and that the expectation values of operators should be the same in both pictures. However it has been shown that they do not produce equivalent results when an actual problem is worked out. Here the initial quantum state $(|\Omega(0)\rangle, \hat{\psi}_S(\vec{x}))$ evolves forward in time under the action of the electric potential given by (32) and (36). In the Heisenberg picture this evolves into the state $(|\Omega(0)\rangle, \hat{\psi}(\vec{x}, t))$. It was shown that it is possible for the expectation value of $\hat{H}_0$ of this quantum state to be negative. If the same problem is examined in the Schrödinger picture the initial state evolves into $(|\Omega(t)\rangle, \hat{\psi}_S(\vec{x}))$. As has been shown (see Eq. (31) the expectation value of $\hat{H}_0$ for this state must be non-negative. Therefore the Heisenberg and Schrödinger pictures do not yield that same result when the expectation value of $\hat{H}_0$ is calculated. This result is consistent with previous research [3] and confirms the comments by Dirac that the Heisenberg picture and Schrödinger picture are not equivalent [4].

## **Appendix**

We will formally show that,

$$H_{0,eH}(t) = H_{0,eS}(t) \tag{54}$$

Start with (34) and convert to the Heisenberg picture by using (13) and (12) to obtain,

$$H_{0,eS}(t) = \langle \Omega(0) | \hat{U}^{-1}(t) \hat{H}_0(\hat{\psi}_S) \hat{U}(t) | \Omega(0) \rangle \tag{55}$$

Next use $\hat{U}(t)\hat{U}^{-1}(t)=1$ and refer to (6) and use the fact that $\hat{U}(t)$ commutes with $H_0$ to obtain,

$$\hat{U}^{-1}(t)\hat{H}_0(\hat{\psi}_S)\hat{U}(t) = \frac{1}{2}\int\left[\hat{U}^{-1}(t)\hat{\psi}_S^{\dagger}\hat{U}(t), H_0\hat{U}^{-1}(t)\hat{\psi}_S\hat{U}(t)\right]d\vec{x} - \xi_r \qquad (56)$$
$$= \hat{H}_0\left(\hat{U}^{-1}(t)\hat{\psi}_S\hat{U}(t)\right)$$

Next prove that,

$$\hat{\psi}(\vec{x},t) = \hat{U}^{-1}(t)\hat{\psi}_S\hat{U}(t) \qquad (57)$$

To show this take the time derivative of $\hat{U}^{-1}(t)\hat{\psi}_S\hat{U}(t)$ to yield,

$$\frac{\partial\left(\hat{U}^{-1}(t)\hat{\psi}_S\hat{U}(t)\right)}{\partial t} = i\begin{pmatrix}\hat{U}^{-1}(t)\hat{H}(\hat{\psi}_S)\hat{\psi}_S\hat{U}(t) \\ -\hat{U}^{-1}(t)\hat{\psi}_S\hat{H}(\hat{\psi}_S)\hat{U}(t)\end{pmatrix} \qquad (58)$$

Use (57) and the fact that $\hat{U}(t)\hat{U}^{-1}(t)=1$ to obtain,

$$\frac{\partial\left(\hat{U}^{-1}(t)\hat{\psi}_S\hat{U}(t)\right)}{\partial t} = i\begin{pmatrix}\hat{H}\left(\hat{U}^{-1}(t)\hat{\psi}_S\hat{U}(t)\right)\hat{U}^{-1}(t)\hat{\psi}_S\hat{U}(t) \\ -\hat{U}^{-1}(t)\hat{\psi}_S\hat{U}(t)\hat{H}\left(\hat{U}^{-1}(t)\hat{\psi}_S\hat{U}(t)\right)\end{pmatrix} \qquad (59)$$

where we have used the relationship $\hat{U}^{-1}(t)\hat{H}(\hat{\psi}_S)\hat{U}(t) = \hat{H}\left(\hat{U}^{-1}(t)\hat{\psi}_S\hat{U}(t)\right)$.

Compare this to (8) to show that $\hat{U}^{-1}(t)\hat{\psi}_S\hat{U}(t)$ and $\hat{\psi}(\vec{x},t)$ obey the same differential equation. Also, due to the fact that $\hat{U}(0)=1$ they are equal at the initial time t=0. Therefore equation (57) is true. Now use (57) in (56) and (55) to obtain $H_{0,eS}(t) = \langle\Omega(0)|\hat{H}_0(\hat{\psi}(\vec{x},t))|\Omega(0)\rangle$. Compare this to (33) to show that (54) is true.